\documentstyle[vsolj01,graphicx,natbib]{article}

\RequirePackage[T1]{fontenc}

\def\cite{\citealt}

\begin{document}

\title{MGAB-V859 and ZTF18abgjsdg: ER UMa-type dwarf novae showing standstills}

\author{Taichi Kato$^1$ and Naoto Kojiguchi$^1$}
\author{$^1$ Department of Astronomy, Kyoto University,
       Sakyo-ku, Kyoto 606-8502, Japan}
\email{tkato@kusastro.kyoto-u.ac.jp}

\begin{abstract}
Using Public Data Release of Zwicky Transient Facility
observations, we found that MGAB-V859 and ZTF18abgjsdg
are dwarf novae and show both ER UMa-type and Z Cam-type states.
There had been only two dwarf novae showing similar
transitions between the ER UMa-type and Z Cam-type states.
MGAB-V859 showed both a transition from the ER UMa-type state
to a long standstill in 2019 and a Z Cam-type state in 2020.
During the standstill in 2020, this object faded twice and
showed dwarf nova-type variations.
ZTF18abgjsdg usually showed ER UMa-type
behavior but standstills were seen in 2018 and 2019.
The supercycles of these objects during the typical
ER UMa-type phase were 55~d and 58~d, respectively.
These objects provide additional evidence that some
ER UMa stars indeed bridge between dwarf nova and novalike
states as proposed in \citet{kat16rzlmi}.
\end{abstract}

\section{Introduction}

   ER UMa stars are a subclass of SU UMa-type dwarf novae
having very short (much less than 100~d) regular supercycles,
very short recurrence times of normal outbursts and long duty cycles
(\cite{kat95eruma}; \cite{rob95eruma}; \cite{pat95v1159ori}; 
for a review, see \cite{kat99erumareview}).
The peculiar behavior of ER UMa stars are believed to be
basically understood by the thermal-tidal instability (TTI) theory
\citep{osa89suuma} under the condition of exceptionally
high mass-transfer rates ($\dot{M}$) \citep{osa95eruma}.
In dwarf novae with even higher $\dot{M}$, the disk can become
thermally stable and standstills in Z Cam stars are considered
to be high-$\dot{M}$, thermally stable states.

   There have been four known objects that showed
both SU UMa-type and Z Cam-type or IW And-type (subtype
of Z Cam stars) behavior.  Two of them
were ER UMa stars.
\begin{itemize}
\item
The ER UMa star RZ LMi showed long-lasting superoutbursts
in 2016--2017 and this behavior was considered to be
a state bridging between an ER UMa star and a novalike
(or Z Cam) star \citep{kat16rzlmi}.
\item
The novalike star BK Lyn showed a temporary ER UMa-type
state in 2011--2012 (\cite{kem12bklynsass}; \cite{pat13bklyn}).
This state might have already started in 2005--2006
\citep{Pdot4}.
\item
The SU~UMa star NY Ser, which is a system close to
an ER UMa star but with supercycles not as regular
as in ER UMa stars,
showed standstills in 2018 and superoutbursts
starting from standstills were observed \citep{kat19nyser}.
The discovery of superoutbursts starting from standstills
provided a clue in understanding still poorly understood
IW And-type dwarf novae (\cite{kat19iwandtype};
\cite{kim20iwandmodel}).
\item
BO Cet has recently shown
both IW And-type and SU UMa-type states \citep{kat21bocet},
but this star is unrelated to an ER UMa star.
\end{itemize}

   We found that two new objects MGAB-V859
(19$^{\rm h}$ 50$^{\rm m}$ 07.038$^{\rm s}$
$+$36\deg 40$^\prime$ 17.24$^{\prime\prime}$, \cite{GaiaEDR3},
discovered by Gabriel Murawski) and
ZTF18abgjsdg
(19$^{\rm h}$ 19$^{\rm m}$ 32.396$^{\rm s}$
$+$31\deg 48$^\prime$ 09.91$^{\prime\prime}$, discovered by
\cite{for21ALeRCE})
show both ER UMa-type states and Z Cam-type standstills using
Public Data Release 6 of
the Zwicky Transient Facility \citep{ZTF}
observations\footnote{
   The ZTF data can be obtained from IRSA
$<$https://irsa.ipac.caltech.edu/Missions/ztf.html$>$
using the interface
$<$https://irsa.ipac.caltech.edu/docs/program\_interface/ztf\_api.html$>$
or using a wrapper of the above IRSA API
$<$https://github.com/MickaelRigault/ztfquery$>$.}.
The peculiar behavior of MGAB-V859 (ER UMa star with
a long standstill) had also been suggested
by Gabriel Murawski in vsnet-chat 8510 on
2020 October 30\footnote{
  $<$http://ooruri.kusastro.kyoto-u.ac.jp/mailarchive/vsnet-chat/8510$>$.
}

\section{MGAB-V859}

\begin{figure*}
  \begin{center}
    \includegraphics[width=16cm]{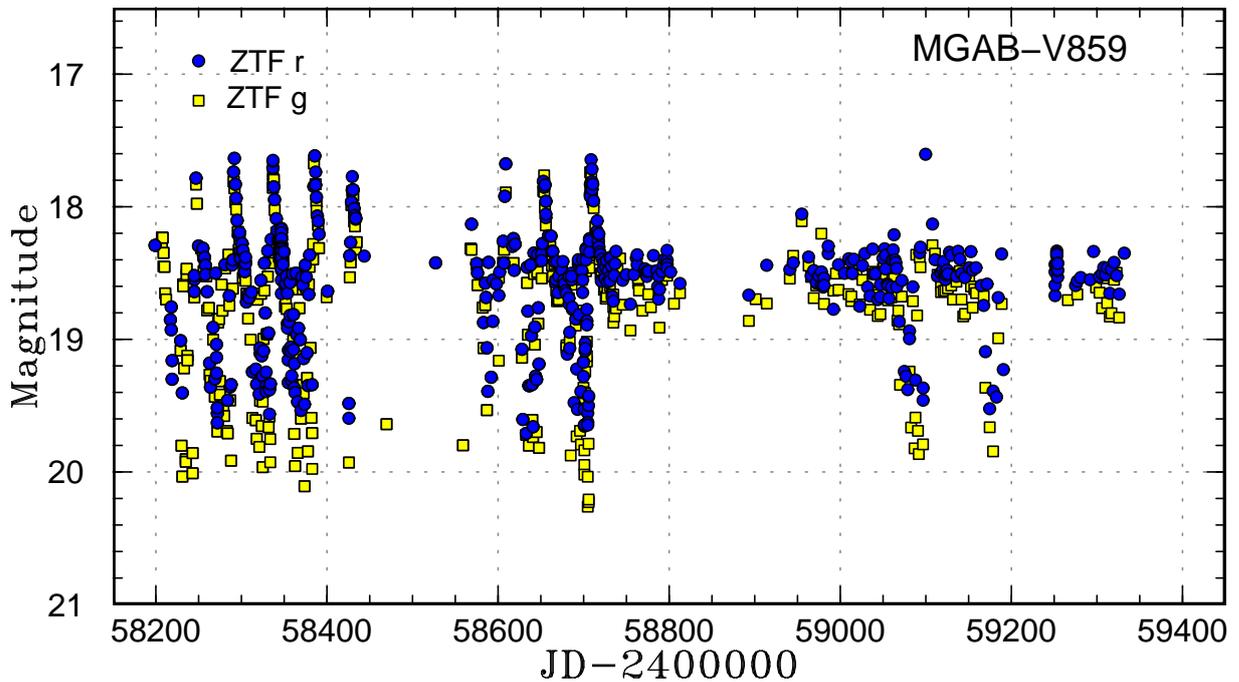}
  \end{center}
  \caption{ZTF light curve of MGAB-V859.  Both ER UMa-type state and
     Z Cam state were present.}
  \label{fig:v859lc}
\end{figure*}

\begin{figure*}
  \begin{center}
    \includegraphics[width=16cm]{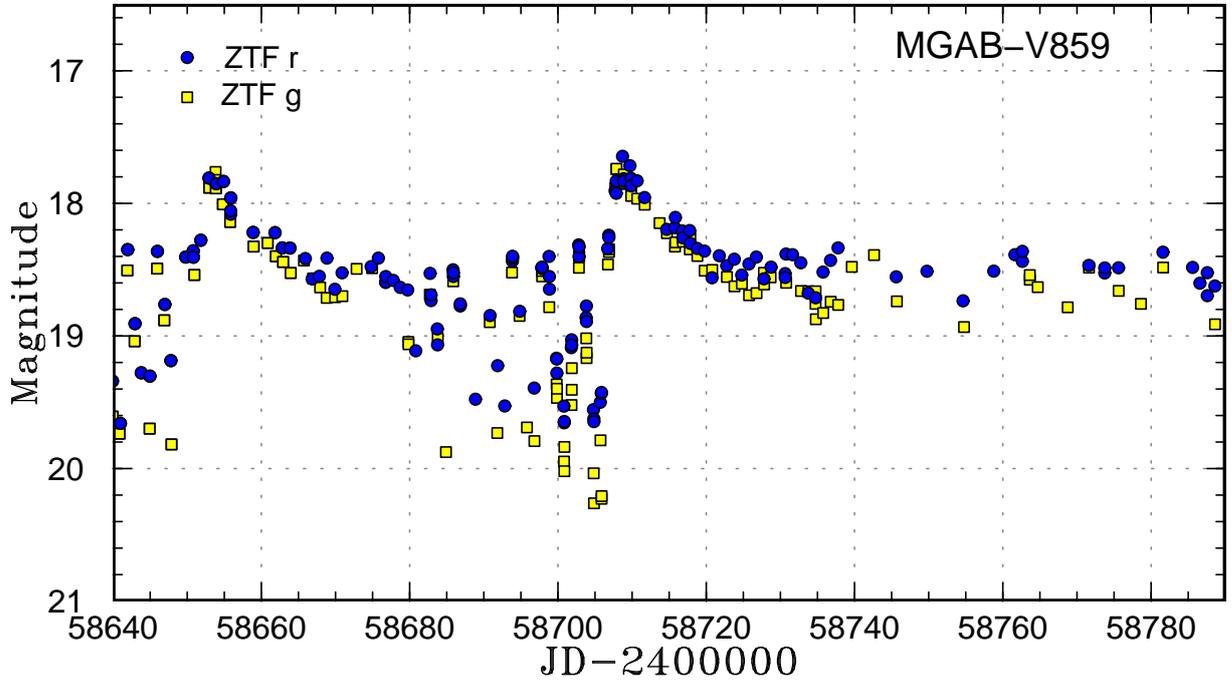}
  \end{center}
  \caption{Enlarged ZTF light curve of MGAB-V859.  The ER UMa-type
     state with a long superoutburst followed by frequent
     normal outbursts (supercycle 55~d) is seen before 
     JD 2458740.  The object smoothly entered a standstill
     following the second superoutburst.}
  \label{fig:v859lc2}
\end{figure*}

\begin{figure*}
  \begin{center}
    \includegraphics[width=16cm]{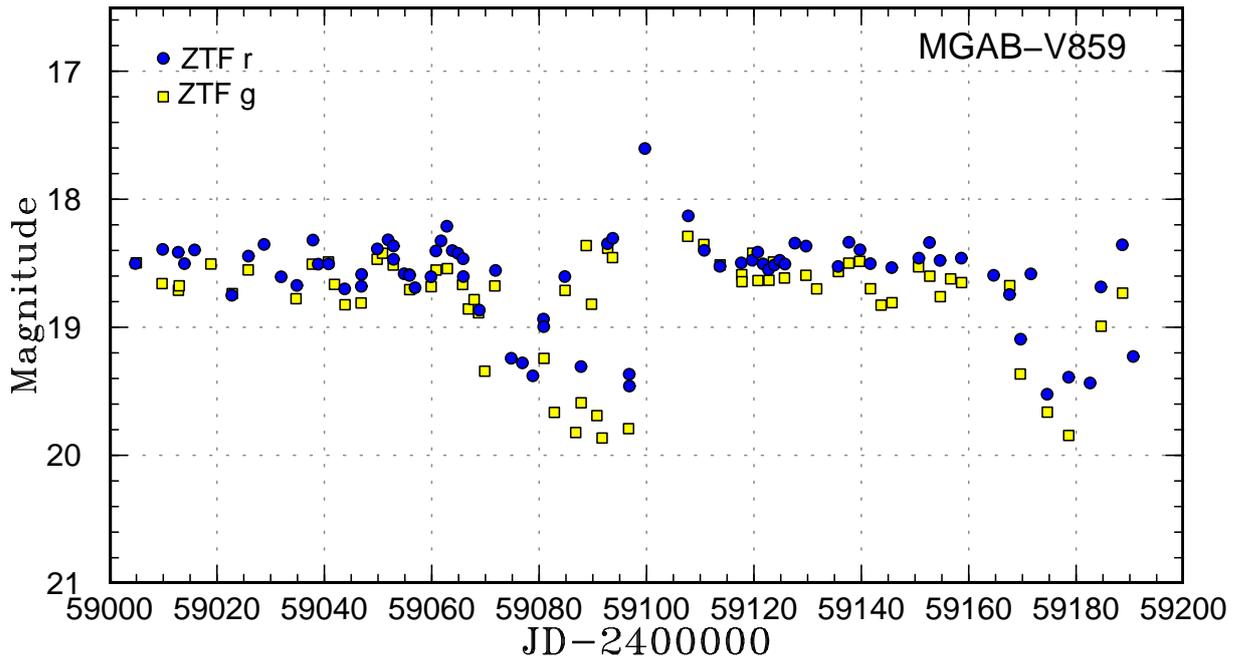}
  \end{center}
  \caption{Enlarged ZTF light curve of MGAB-V859 during
     the Z Cam state in 2020.  Standstills were terminated
     by fading as in ordinary Z Cam stars.}
  \label{fig:v859lc3}
\end{figure*}

   The entire light curve of MGAB-V859 is shown in
figure \ref{fig:v859lc}.  Both ER UMa-type state and
Z Cam-type state were present.  The enlargement of the ER UMa-type
state is shown in figure \ref{fig:v859lc2}.
Short supercycle (55~d), frequent normal outbursts and
the long duty cycle ($\sim$0.5) are characteristic to
an ER UMa star.  Following the second superoutburst,
this object entered a standstill.
Figure \ref{fig:v859lc3} shows an enlargement of
the Z Cam-type state in 2020.  In contrast to NY Ser
\citep{kat19nyser}, this object faded from standstills
just as in ordinary Z Cam stars.  There was a hint of
low-amplitude oscillations with a period of $\sim$10~d
just before the termination of the first standstill.
This behavior seems to indicate that the disk was becoming
thermally unstable before the termination of
the standstill.

\begin{figure*}
  \begin{center}
    \includegraphics[width=13cm]{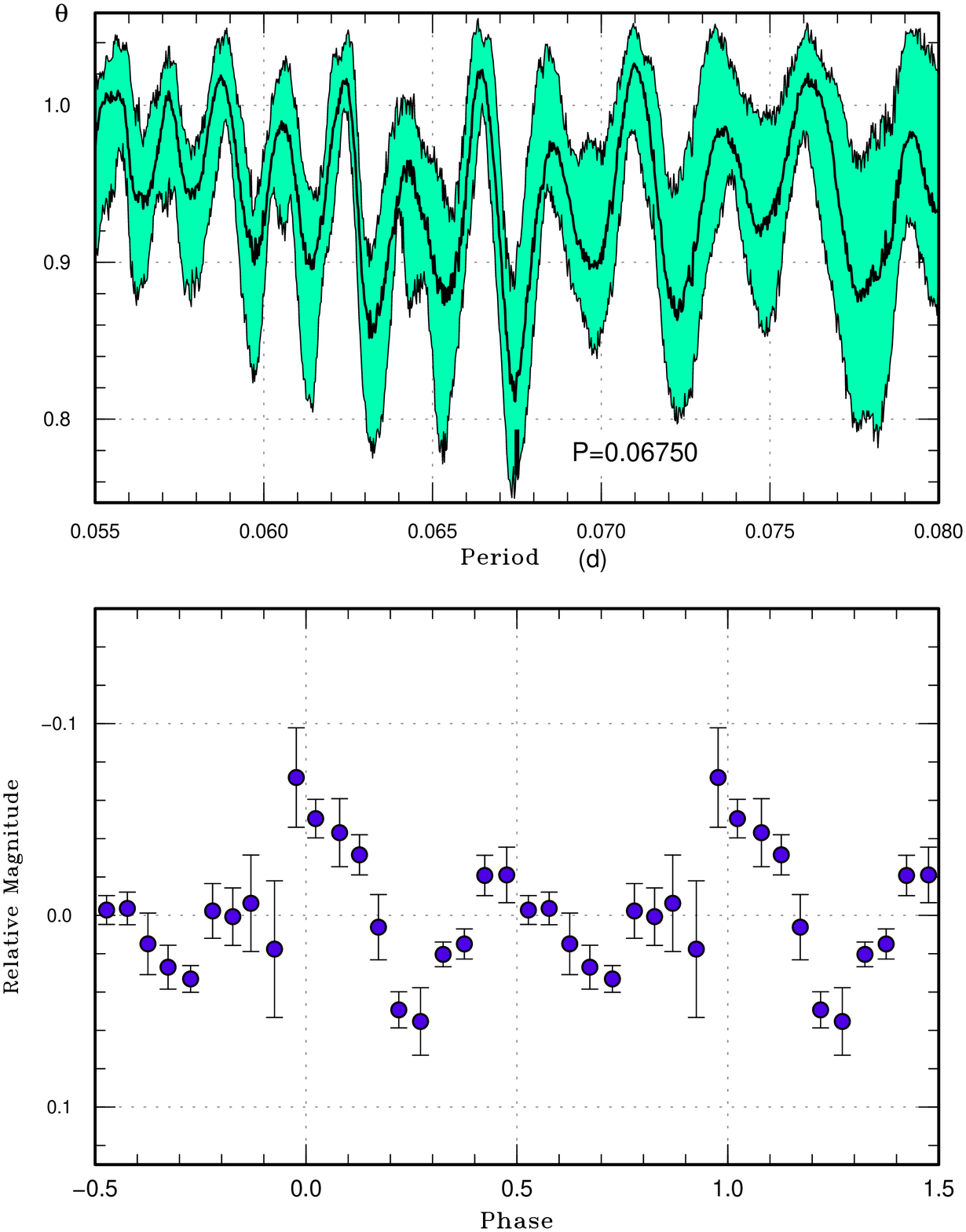}
  \end{center}
  \caption{Possible superhumps in MGAB-V859 during
     a superoutburst from the ZTF time-resolved data.
     (Upper): PDM analysis.  We analyzed 100 samples which
     randomly contain 50\% of observations, and performed
     the PDM analysis for these samples.
     The bootstrap result is shown as a form of 90\% confidence
     intervals in the resultant PDM $\theta$ statistics.
     (Lower): Phase-averaged profile.}
  \label{fig:v859pdm}
\end{figure*}

   There were time-resolved observations in ZTF
between JD 2459345.68 and 2459346.81, 9--10~d after
the peak of the superoutburst.  A period analysis using
Phase Dispersion Minimization (PDM, \cite{PDM}) yielded
a possible superhump period of 0.0675(2)~d
(figure \ref{fig:v859pdm}).  The 1$\sigma$ error for
the PDM analysis was estimated by the methods
of \citet{fer89error} and \citet{Pdot2}.
Since the baseline was very short and superhumps in
ER~UMa stars tend to decay rather quickly
(cf. \cite{kat96erumaSH}), this period needs to be
confirmed by independent observations near the peak
of a superoutburst.

\section{ZTF18abgjsdg}

\begin{figure*}
  \begin{center}
    \includegraphics[width=16cm]{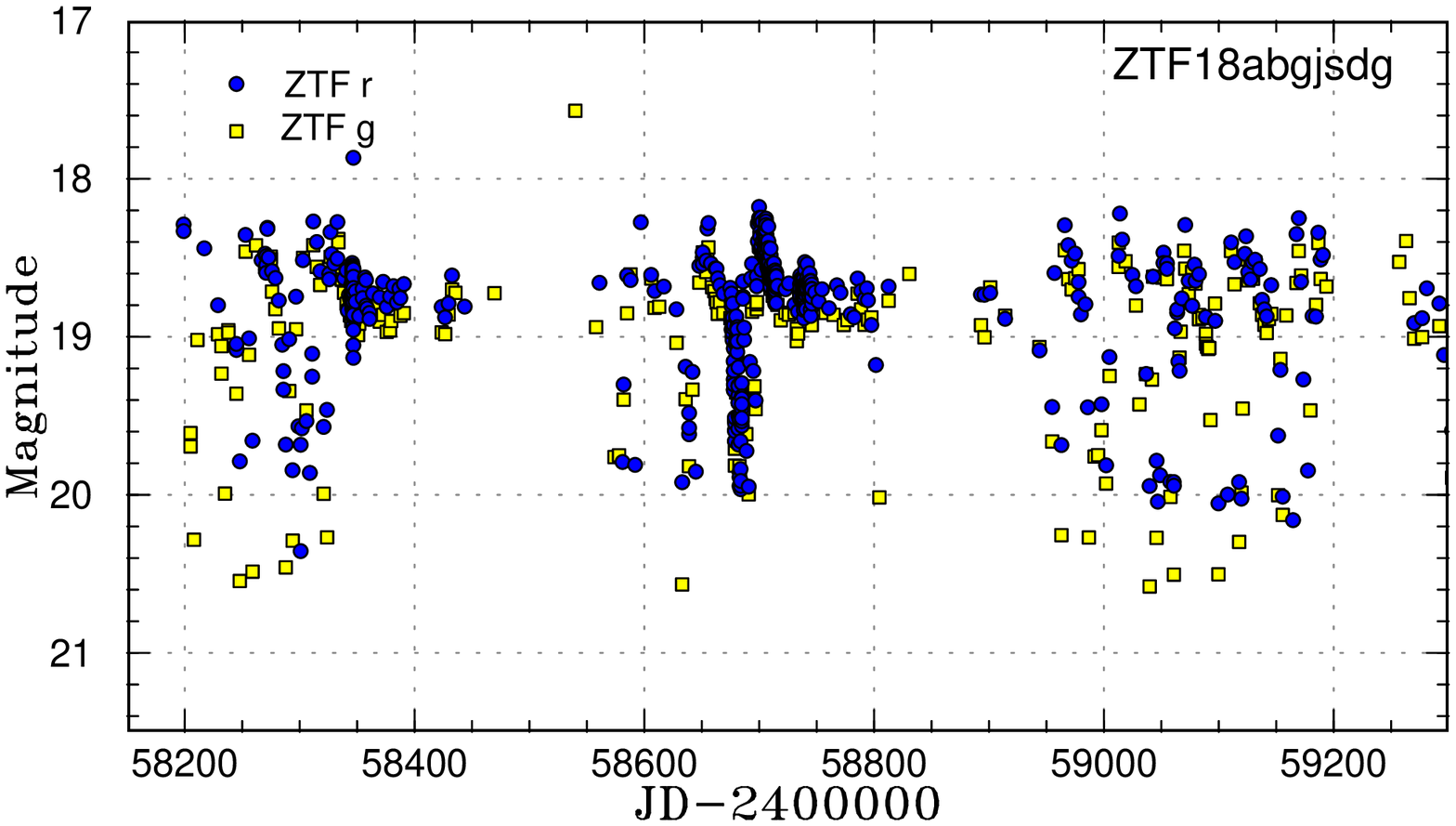}
  \end{center}
  \caption{ZTF light curve of ZTF18abgjsdg. 
     Both ER UMa-type state and Z Cam state were present.}
  \label{fig:ztf18abgjsdg}
\end{figure*}

   The case of ZTF18abgjsdg is shown in
figure \ref{fig:ztf18abgjsdg}.  Although the data were
less abundant than in MGAB-V859, standstills following
superoutbursts on JD 2458326 and 2458699 were recorded.
The supercycles were relatively regular ($\sim$58~d)
after JD 2458966, but were variable JD 2458720.
The variable supercycles were similar to what were
observed in RZ LMi in 2016--2017 \citep{kat16rzlmi}
and probably occurred as a result of changing $\dot{M}$.

   Although these two objects are very faint, time-resolved
observations are encouraged to establish the superhump periods
and to study the relation between superhumps and
ER UMa-Z Cam transitions.

\section*{Acknowledgments}

This work was supported by JSPS KAKENHI Grant Number 21K03616.

Based on observations obtained with the Samuel Oschin 48-inch
Telescope at the Palomar Observatory as part of
the Zwicky Transient Facility project. ZTF is supported by
the National Science Foundation under Grant No. AST-1440341
and a collaboration including Caltech, IPAC, 
the Weizmann Institute for Science, the Oskar Klein Center
at Stockholm University, the University of Maryland,
the University of Washington, Deutsches Elektronen-Synchrotron
and Humboldt University, Los Alamos National Laboratories, 
the TANGO Consortium of Taiwan, the University of 
Wisconsin at Milwaukee, and Lawrence Berkeley National Laboratories.
Operations are conducted by COO, IPAC, and UW.

The ztfquery code was funded by the European Research Council
(ERC) under the European Union's Horizon 2020 research and 
innovation programme (grant agreement n$^{\circ}$759194
-- USNAC, PI: Rigault).

\end{document}